\documentclass[aps,superscriptaddress,showpacs,floatfix,nofootinbib]{revtex4-1}
\usepackage{color}
\usepackage{graphicx}	
\usepackage{amsthm}

\newtheorem*{lemma}{Lemma}
\newtheorem*{dilemma}{Dilemma}

\begin{document}

\title{Do nuclear collisions create a locally equilibrated quark-gluon plasma?}

\author{P.~Romatschke}
\affiliation{Department of Physics, 390 UCB, University of Colorado at Boulder, Boulder, Colorado 80309, USA}
\affiliation{Center for Theory of Quantum Matter, University of Colorado, Boulder, Colorado 80309, USA}

\date{\today}

\begin{abstract}
Experimental results on azimuthal correlations in high energy nuclear collisions (nucleus-nucleus, proton-nucleus and proton-proton) seem to be well described by viscous hydrodynamics. It is often argued that this agreement implies either local thermal equilibrium or at least local isotropy. In this note, I present arguments why this is not the case. Neither local near-equilibrium nor near-isotropy are required in order for hydrodynamics to offer a successful and accurate description of experimental results. However, I predict the breakdown of hydrodynamics at momenta of order seven times the temperature, corresponding to a smallest possible QCD liquid drop size of \hbox{0.15 fm}.
\end{abstract}

\maketitle

\tableofcontents

\section{Preface}

The intention of this note is to provide an impulse to the nuclear collisions community which in my opinion is trying to reconcile the traditional paradigm of hydrodynamic applicability with the “unreasonable success” of hydrodynamics at seemingly describing ever smaller systems. In an attempt of thinking ``outside-the-box'', my hope is to start a discussion as regards to the interpretation of this hydrodynamic success that challenges the traditional paradigm of local equilibration. Aiming at challenging the traditional, I found it very hard to conform to the standard manuscript style, which led to this informal note. 

It should be pointed out that many of my main points have been made by others before me. All I have done is collect the available information, and based on this information, offer my own interpretations, conclusions and predictions. I expect that these conclusions may appear obvious to some readers and contentious to others. I invite readers who do not agree with my statements to engage in a constructive dialogue, which I hope could help us make progress.

\section{Introduction}

If nuclear collision experiments do not probe near-equilibrium matter, then this may have a number of consequences which to my knowledge have not been appreciated before, providing the motivation for this note. Firstly, it would imply that the nuclear experiments do not (directly) probe equilibrium QCD properties as those calculated in first-principle lattice QCD calculations. Depending on the degree of non-equilibrium, experiments may be closer or farther away from the “QCD phase diagram plane” spanned by temperature and baryon chemical potential. While for hydrodynamics, a projection from “non-equilibrium space” to the equilibrium plane is provided by e.g. the Landau matching condition, for other observables such a projection is not explicitly known. For instance, it is possible that the phenomenon of critical fluctuations associated with the experimental search for a QCD critical point would get modified when experiments probe QCD away from equilibrium. 

The understanding that the matter created in high energy nuclear collisions does not need to equilibrate or isotropize locally in order for hydrodynamics to be quantitatively applicable would imply that the ``early thermalization puzzle'' is to some extent not a genuine puzzle (see more details about this below). On the other hand, having experimental access to a non-equilibrium quantum system could lead to new directions in the field such as e.g. observing non-equilibrium entropy production, properties of non-thermal fixed points or off-equilibrium photo-production.

Finally, if systems created in nuclear collisions do not equilibrate this could naturally explain why proton-proton collision data on azimuthal correlations appears to be so similar to data obtained in nucleus-nucleus collisions. Once small gradients or near-equilibrium is no longer a requirement, hydrodynamics will generically convert initial state geometry and fluctuations into correlations, thus making large and small systems look alike in their azimuthal correlation signals. Pushing this idea even further would imply that any “lump” of sufficiently high energy density could expand according to the laws of hydrodynamics (with one important caveat which will be discussed below). A natural consequence of this would be the presence of exponentially falling (“thermal”) spectra as well as potential azimuthal correlations in $e^+$+$e^-$ collisions.

\section{A historical perspective}

In recent years, it has been demonstrated that experimental results obtained in relativistic nuclear collisions are well described by hydrodynamic simulations. Based on the paradigm that  hydrodynamics requires near-equilibrium in order to be applicable, this successful hydrodynamic description has been interpreted as evidence for a locally equilibrated state of matter (dubbed the ‘quark-gluon plasma’) in high energy nuclear collisions.

The question on how the system created in high energy nuclear collisions reaches or at least comes close to equilibrium subsequently has led to a number of developments. In particular, it was realized that because of the expansion of the matter into the vacuum following the nuclear collision, the system would cool and thus freeze into a hadronic gas quickly, at which point the type of correlations observed in experiment could not longer be built up. Thus, it became apparent that a fluid dynamic approximation to the system dynamics had to start early, on a time-scale of $\tau\sim 1$ fm/c or less \cite{Heinz:2001xi,Luzum:2008cw,Schenke:2010nt,Niemi:2015qia}.

In a seminal article entitled “Bottom-up Thermalization” \cite{Baier:2000sb}, Baier, Mueller, Schiff and Son calculated the time when gluons at weak coupling $\alpha_s\ll 1$ reached thermal equilibrium in a heavy-ion collision, finding \cite{Baier:2002bt}
\begin{equation}
\tau\geq 1.5 \alpha_s^{-13/5} Q_s^{-1}
\end{equation}
which leads to $\tau\gtrsim 6.9$ fm/c for $Q_s\sim 1$ GeV and $\alpha_s\sim 0.3$. Clearly, this result seemed to be in tension with the starting time required from the agreement between hydrodynamics and experimental data.

Arnold, Lenaghan, Moore and Yaffe \cite{Arnold:2004ti} pointed out that a possible way out of the dilemma was that full thermalization was actually not required for a hydrodynamic description, and that local (near-) isotropy of the pressure tensor was sufficient. Thus, the attention of the field shifted towards finding a mechanism to quickly achieve local isotropy (“isotropization”) rather than full thermalization in high energy nuclear collisions.

One such possible mechanism was that of non-abelian plasma instabilities, specifically the non-abelian Weibel instability, which had been extensively studied by Mrowczynski since the 1980s \cite{Mrowczynski:1988dz,Mrowczynski:1993qm}. In a series of numerical studies by a number of groups the growth and saturation of these plasma instabilities was determined for non-expanding systems \cite{Rebhan:2004ur,Arnold:2005vb,Dumitru:2005gp,Bodeker:2007fw,Berges:2007re}, see Refs.~\cite{Mrowczynski:2016etf,Fukushima:2016xgg} for a review. While corroborating the initial exponential approach towards isotropy, these numerical studies suggested the system to stall at large pressure anisotropies once the plasma instabilities reached the non-perturbative non-abelian scale and could no longer grow exponentially. Even worse, later studies in expanding systems aiming at more realistically describing experimental nuclear collisions indicated that the effect of plasma instabilities was delayed/diminished to an extent that they could not lead to local pressure isotropy in a time-scale relevant for nuclear collisions at RHIC and the LHC \cite{Romatschke:2006nk,Romatschke:2006wg,Rebhan:2008uj,Berges:2012iw}.

While full isotropization seemed difficult to achieve within a weak-coupling QCD based framework, it appeared to be reachable much faster in gravitational duals of gauge theories in the limit of infinite coupling. For instance, Chesler and Yaffe \cite{Chesler:2008hg} report isotropization to occur at $\tau\sim 0.7/T$ for a non-expanding system, roughly translating to $\tau\simeq 0.35$ fm/c when assuming $T\sim 0.4$ GeV. However, similar to the case of plasma instabilities, isotropization does take more time when considering the case of expanding systems such as those in nuclear collisions, because expansion constantly tries to drive the system away from local isotropy. This is the reason why in newer studies including expansion \cite{Chesler:2010bi,Casalderrey-Solana:2013aba}, the isotropization time gets delayed. In particular, it eventually became clear that even for systems at infinite coupling strength the system does not isotropize early. Rather, even at infinite coupling, the pressure anisotropy exceeds 10 percent for all times $\tau\lesssim 10$ fm/c \cite{Keegan:2015avk}.

For completeness, it should be noted that when including inelastic processes in a weak-coupling based descriptions, recent studies \cite{Kurkela:2011ti,Gelis:2013rba} have demonstrated the approach to isotropy, albeit at times later than those found for infinitely strongly coupled gauge theories. (This better be the case). Thus, the approach to isotropization in expanding gauge theories is now understood both at weak and strong coupling, and indicates long times.

Despite the impressive progress made, I believe it is a correct statement to say that at phenomenologically relevant times of $\tau\sim 1$ fm/c following a nuclear collision, \textit{no} theoretical approach (be it weakly coupled or strongly coupled) finds the longitudinal and transverse pressure to agree with each other to better than a factor of two. Obviously, a pressure anisotropy of 50 percent is not close to an “isotropic” system, let alone a system in thermal equilibrium. By the criterion of Arnold, Lenaghan, Moore and Yaffe, hydrodynamics should not apply. 

But it does.

\newpage
\section{Hydrodynamization or The Onset of Hydrodynamic Applicability}

\begin{itemize}
\item
Q: How do you people know hydrodynamics applies for pressure anisotropies of 50 percent or larger? 
\item
A: We checked.
\end{itemize}

\begin{figure}[t]
\includegraphics[width=0.7\linewidth]{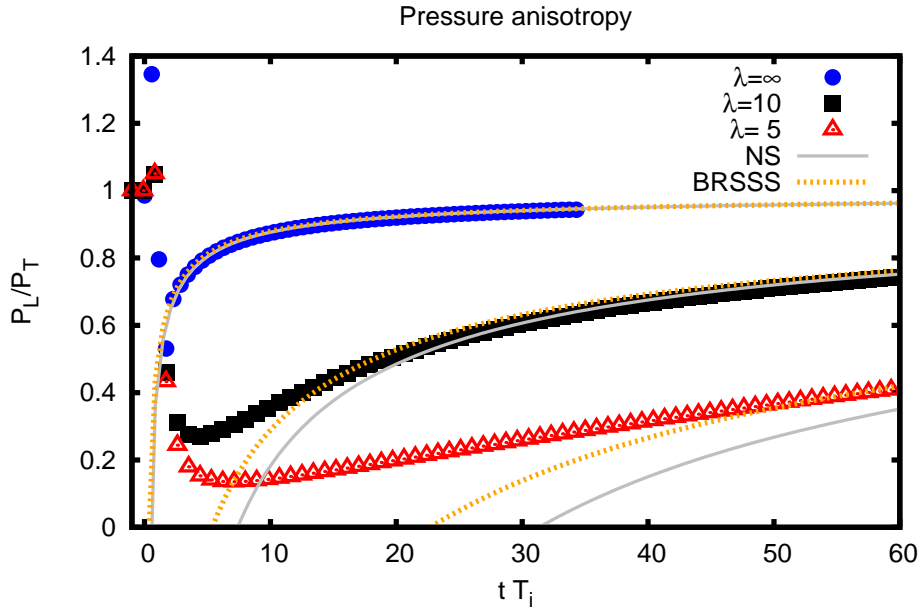}
\caption{Pressure anisotropy versus time for matter undergoing Bjorken-like expansion. Shown are exact results for matter with different coupling constant $\lambda$ (symbols, calculated using AdS/CFT and kinetic theory), as well as hydrodynamics in first and second order gradient expansion ('NS' and 'BRSSS', respectively). Note that $P_L/P_T=1$ would correspond to isotropy (ideal hydrodynamics). Figure adapted from Ref.~\cite{Keegan:2015avk}.
\label{fig:one}}
\end{figure}

Let us consider the following numerical experiment. Take matter described by gauge/gravity duality at infinite coupling or alternatively described by kinetic theory at some finite (constant) value of the coupling. Let the matter be initially at rest in equilibrium with some temperature $T_i$ in flat Minkowski space. Then, at a time $t\sim 0$, the spacetime suddenly starts to expand in one dimension so that it effectively mimics the effects of so-called Bjorken flow \cite{Bjorken:1982qr}. The symbols in Fig.~\ref{fig:one} show the response of the matter (at various values of the coupling $\lambda$) in terms of the ratio of longitudinal to transverse pressure as a function of time. The matter is initially in equilibrium so $P_T=P_L$ (zero pressure anisotropy) and also tends to equilibrium at late times when the expansion becomes very slow. However, for $t\simeq 0$ when the expansion is most rapid, the matter is clearly not locally isotropic, and deviations from local isotropy become larger as the coupling is decreased.

Also shown in Fig.~\ref{fig:one} are results from a hydrodynamic gradient expansion to first and second order in gradients, respectively. One observes that hydrodynamics quantitatively matches the exact results whenever the pressure anisotropy is 50 percent or smaller. 

This 'unreasonable success' of hydrodynamics in describing systems with pressure anisotropies of order unity is neither limited to this one example nor to AdS/CFT dynamics, nor exclusively to previous work by the present author, cf. Ref.~\cite{Chesler:2009cy,Heller:2011ju,Wu:2011yd,vanderSchee:2012qj,Casalderrey-Solana:2013aba,Kurkela:2015qoa}.

While of course no general proof, the above numerical experiment indicates that hydrodynamics is able to give quantitatively accurate descriptions even when the matter not locally isotropic. The timescale at which hydrodynamics first is able to closely approximate the subsequent dynamics of the “exact” underlying microscopic theory has been dubbed “hydrodynamization time” \cite{CasalderreySolana:2011us}. At the hydrodynamization time, the matter is typically not locally isotropic. So what sets the timescale for the onset of the applicability of hydrodynamics?

\section{Hydrodynamic versus Non-Hydrodynamic Modes}
\label{sec:this}

What is hydrodynamics? The equations of hydrodynamics can be derived using a multitude of approaches. Some assume the system to be close to thermal equilibrium, others assume a weakly coupled microscopic particle description (kinetic theory). In my opinion, the most general derivation of hydrodynamics follows the approach of effective field theory (EFT). 

According to this viewpoint, hydrodynamics is the EFT of long-lived, long-wavelength excitations consistent with the basic symmetries of the underlying system. The fundamental variables of the EFT are that of a fluid: pressure P, (energy) density $\epsilon$ and fluid velocity $u^a$. To lowest (leading) order in the EFT, only algebraic combinations of these quantities will enter the description\footnote{As one important qualifier, let me point out that a necessary condition for this to work is the presence of a local rest frame, cf. the discussion in Ref.~\cite{Arnold:2014jva}. Without a local rest frame, the local energy density cannot be defined, and a fluid EFT approach is not applicable.}. Corrections can then be systematically obtained by considering gradients of the fundamental variables. 

Applying this EFT approach to the energy-momentum tensor for a relativistic system in three dimensions leads to the well-known expansion
\begin{equation}
\label{Tns}
T^{ab}=(\epsilon+P) u^a u^b+P g^{ab}-2 \eta \nabla^{\langle a} u^{b \rangle}+\ldots\,,
\end{equation}
where $g^{ab}$ denotes the space-time metric, $\eta$ is the shear viscosity coefficient and the symbols $<>$ denote a particular symmetric projector that dedicated readers can look up e.g. in Ref.~\cite{Baier:2007ix}.
With Eq.~(\ref{Tns}), conservation of energy and momentum $\nabla_a T^{ab}=0$  then {\it are} the relativistic Navier-Stokes equations. Many articles have been written about non-causality of the relativistic Navier-Stokes equations; I will simply ignore this issue here because it is somewhat tangential to the following discussion. 

The above EFT derivation does at no point invoke the presence of an underlying particle-based, kinetic description of the matter. However, given the requirement of the small gradients, it does seem to require the system to be close to isotropy.
So what if gradients were not small in a particular situation of interest? Obviously, stopping at first order in a gradient expansion would not be a good approximation. However, one could try to include higher order gradient corrections to obtain a good approximation. I will try to elucidate what happens in this case through a particular example.

For pedagogical purposes, let me pick the example of ${\cal N}=4$ SYM at infinite coupling undergoing Bjorken expansion that has been worked through in a tour-de-force paper by Heller, Janik and Witaszczyk \cite{Heller:2013fn}. In this case one has a high degree of symmetry, and the only relevant gradient is $\nabla\cdot u=\frac{1}{\tau}$. The equations of motion then lead to a solution for the temperature $T$ as a function of $\tau$ which may be systematically calculated for small gradients (or equivalently large $\tau$). (Note that the actual dimensionless expansion parameter is $\frac{1}{\tau T}$ which scales as $\tau^{-2/3}$ in the hydrodynamic limit).  Calculating the temperature $T(\tau)$ in a hydrodynamic gradient expansion to order 240 leads to a series of the form \cite{Heller:2013fn}
\begin{equation}
\label{eq:Tnaive}
T(\tau)=\hat \tau^{-1/3}\left(1+\sum_{n=1}^{240} \alpha_n  \hat \tau^{-2n/3}\right)\,,
\end{equation} 
where $\hat \tau=\frac{\tau}{\tau_0}$, $\alpha_n$ constant, and $\tau_0$ setting the initial time (or equivalently temperature) scale.

If the gradient expansion was convergent, then we would have succeeded in a (high order) theory of hydrodynamics that was unconditionally applicable also when the gradients are large. Given that for this theory a very large number of coefficients $\alpha_n$ had to be calculated, it would be cumbersome if not impossible to generalize this approach to situations with a much lower degree of symmetry (e.g. nuclear collisions), but at least in principle, it would work!

Unfortunately, there is mounting evidence that the hydrodynamic gradient expansion generally is not a convergent series. In the cases that have been examined in detail (${\cal N}=4$ and ${\cal N}=2^*$ SYM at infinite coupling, weakly coupled kinetic theory in the relaxation time approximation and  M\"uller-Israel-Stewart (MIS) theory) it was found that $\alpha_n\propto n!$ for large n, thus making the gradient expansion a divergent series \cite{Heller:2013fn,Heller:2015dha,Buchel:2016cbj,Denicol:2016bjh,Heller:2016rtz}.

However, not all is lost. It turns out that when inspecting the analytic structure of gradient expansions at high orders, it is possible to use a generalized Borel resummation to rewrite the above series for the case of ${\cal N}=4$ SYM as
\begin{equation}
\label{eq:Tres}
T(\tau)=T_{\rm hydro}(\tau)
+\gamma \exp{\left[-i \int d\hat\tau \left(\hat{\omega}_{\rm Borel} \hat\tau^{-1/3}+\sum_{n=1} \hat\omega_{n}\hat\tau^{-(2n+1)/3}\right]\right)}+\ldots\,,
\end{equation}
where $T_{\rm hydro}(\tau)$ is well approximated by the first few orders in (\ref{eq:Tnaive}) as long as $\hat\tau$ is not too small. In the above expression, $\hat{\omega}_{\rm Borel}\simeq\pm 3.1193-2.7471 i$, and both $\gamma,\omega_1$ have been calculated in Ref.~\cite{Heller:2013fn}.

There are two things to note about the resummed result (\ref{eq:Tres}). First, the exponential multiplying the coefficient $\gamma$ in (\ref{eq:Tres}) can not be recast in terms of the hydrodynamic gradient expansion. It is a truly non-hydrodynamic mode, and its presence explains why the naive hydrodynamic gradient series is divergent. 

Second, the numerical value of $\hat{\omega}_{\rm Borel}$ is not an arbitrary number. It happens to be consistent with the first non-hydrodynamic quasi-normal mode frequency of a near-equilibrium 5d Schwarzschild-AdS black hole
\begin{equation}
\hat{\omega}_{\rm QNM}^{(1)}=\pm 3.119452-2.746676 i\,,
\end{equation}
calculated by Starinets in Ref.~\cite{Starinets:2002br}. A quasi-normal mode corresponds to a pole of a two-point function in the complex frequency plane located at $\omega=2\pi T \hat{\omega}_{\rm QNM}$. Linear response then implies the presence of a contribution of the form $e^{i \omega \tau}=e^{i 2\pi \hat{\omega} T \tau}$ to the one-point function which upon using the leading order expression in (\ref{eq:Tnaive}) for $T(\tau)$ then leads to a result $e^{\sim i \hat{\omega} \hat\tau^{2/3}}$ consistent with (\ref{eq:Tres}), cf.~\cite{Janik:2006gp}. While only the first non-hydrodynamic quasi-normal mode $\hat{\omega}_{\rm QNM}^{(1)}$ has been identified in the Borel transform of the gradient expansion, it is likely that all higher non-hydrodynamic modes also will contribute likewise to $T(\tau)$, which has been anticipated through the ellipses in (\ref{eq:Tres}). In fact, Buchel, Heller and Noronha were able to show that for the case of ${\cal N}=2^*$ SYM the first 10 quasi-normal modes could be obtained from the relevant Borel transform \cite{Buchel:2016cbj}.

Thus, the following picture emerges: a naive hydrodynamic gradient expansion of the energy-momentum tensor is divergent because of the presence of other, non-hydrodynamic degrees of freedom. However, the contribution from these non-hydrodynamic modes may be either resummed via a generalized Borel transform, or anticipated through explicitly calculating the non-hydrodynamic mode structure of the energy-momentum tensor for the theory under consideration:
\begin{equation}
T^{ab}=T^{ab}_{\rm hydro}+T^{ab}_{\rm non-hydro}\,.
\end{equation}

The term $T_{\rm hydro}(\tau)$ in (\ref{eq:Tres}) is a ``generalized'' hydrodynamic piece has been dubbed ``hydrodynamic attractor'' or ``all order hydrodynamics'' by various authors \cite{Bu:2014ena,Heller:2015dha}. As remarked above, it is well approximated by a low-order gradient expansion approximation even in regime when gradients are moderately strong (see e.g. Ref.~\cite{Heller:2015dha}). Thus, even though different in principle, it is entirely conceivable that for many applications involving gradient terms of order unity, $T^{ab}_{\rm hydro}$ will quantitatively be well approximated by the naive, low-order gradient expansion (\ref{Tns}) because the divergence of the gradient approximation only becomes apparent when including higher orders. High-temperature perturbation theory in QCD exhibits a similar feature where the leading order $(g^2)$ correction to the free energy offers a quantitatively reasonable description of lattice QCD data even for $g\simeq 1$ \cite{Blaizot:2003iq}, while the inclusion of higher order corrections clearly exhibits the divergent series nature of a naive perturbative expansion.

Contrary to the hydrodynamic contribution, the non-hydrodynamic contribution $T^{ab}_{\rm non-hydro}$ will in general not have a universal form, but rather be dependent on the particular underlying microscopic description under consideration (``microscopic'' in the sense of QCD, not in the sense of quasi-particles) .

This is most easily elucidated when considering the small amplitude (but arbitrary gradient) linear response of the energy-momentum tensor to an initial source $S^{cd}({\bf x})$,
\begin{equation}
\label{eq:delT}
\delta T^{ab}(t,{\bf x})=\int d\omega d^3k e^{-i \omega t+i {\bf k}\cdot {\bf x}} G_R^{ab,cd}(\omega,{\bf k}) S^{cd}({\bf k})\,,
\end{equation}
where $G_R^{ab,cd}(\omega,{\bf k})$ is the retarded two-point function of the energy-momentum tensor (cf.~\cite{Kovtun:2012rj}). 

\begin{figure}[t]
\includegraphics[width=0.7\linewidth]{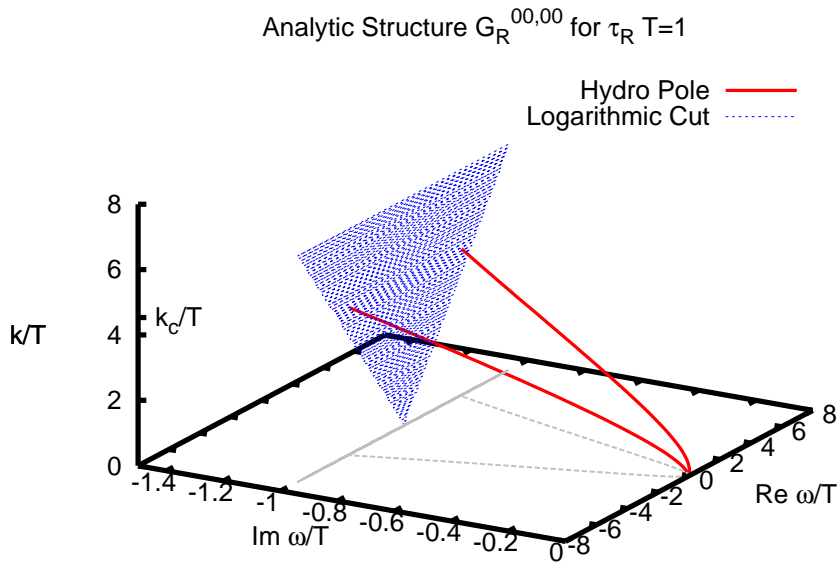}
\caption{Analytic structure of the $G_R^{00,00}(\omega, {\bf k})$ correlator in kinetic theory (from Ref.~\cite{Romatschke:2015gic}). At low $k/T$, there are two hydrodynamic poles and a logarithmic branch cut from $-k-\frac{i}{\tau_R}$ to $k-\frac{i}{\tau_R}$. Increasing $k/T$, the hydrodynamic poles acquire larger and larger negative imaginary part until at $k_c/T\simeq 4.53$ they merge with the branch cut and disappear from the physical Riemann sheet. For $k>k_c$, only the branch cut remains. Grey lines are projections of trajectories to the $k=0$ plane.
\label{fig:two}}
\end{figure}

For instance, in the case of ${\cal N}=4$ SYM at infinite coupling, $G^{00,00}(\omega,{\bf k})$ possesses only poles in the complex plane. Two of these poles may be uniquely identified as hydrodynamic sound poles, located at $\omega_h=\pm c_s |{\bf k}| -\frac{2 i \eta |{\bf k}|^2}{3 s}$ when $|{\bf k}|\ll 1$. In addition to the hydrodynamic poles, there is an infinite number of (pairs) of non-hydrodynamic quasinormal modes located at $\omega=\omega_{nh}^{(1)},\omega_{nh}^{(2)},\ldots$ \cite{Kovtun:2005ev}. Performing the frequency integration in (\ref{eq:delT}) will pick up contributions at all of these poles, immediately leading to
\begin{equation}
\label{eq:delT2}
\delta T^{00}(t,{\bf x})=\int d^3k e^{i {\bf k}\cdot {\bf x}} \left[e^{-i\omega_h t}a_h({\bf k})+\sum_{n=1}^\infty e^{-i\omega_{nh}^{(n)} t}a_n({\bf k)}\right]=\delta T^{00}_{\rm hydro}+\delta T^{00}_{\rm non-hydro}\,,
\end{equation}
where the coefficient functions $a_h({\bf k}),a_n({\bf k})$ depend on the residues from the integration as well as the source function $S({\bf k})$. The integral in (\ref{eq:delT2}) cleanly separates into a hydrodynamic piece governed by the sound mode dispersion $\omega_h(k)$ and an infinite sum over non-hydrodynamic contributions with dispersion $\omega_{nh}^{(n)}(k)$.

More important than the realization that the energy-momentum tensor can be split into a hydrodynamic and a non-hydrodynamic piece is the fact that the hydrodynamic poles cease to exist at some value of ${\bf k}$ when the coupling is finite \cite{Grozdanov:2016vgg}. This implies that for $|{\bf k}|$ larger than some critical value of $|{\bf k}|_c$ (dependent on the coupling), the hydrodynamic component vanishes from the spectrum and is replaced by purely non-hydrodynamic behavior. At least for $|{\bf k}|>|{\bf k}|_c$, hydrodynamics has broken down.

One may criticize that ${\cal N}=4$ SYM is a very special microscopic theory, and worry about drawing general conclusions based exclusively on ${\cal N}=4$ SYM. However, it turns out that when calculating $G_R^{ab,cd}(\omega,{\bf k})$ in kinetic theory in the relaxation time approximation \cite{Romatschke:2015gic}, similar conclusions apply. In kinetic theory, $G_R^{00,00}(\omega,{\bf k})$ generally has two hydrodynamic poles which are located at $\omega_h=\pm c_s |{\bf k}| -\frac{2 i \eta |{\bf k}|^2}{3 s}$ when $|{\bf k}|\ll 1$. In addition to these hydrodynamic poles, $G_R^{00,00}(\omega,{\bf k})$ exhibits a logarithmic branch cut which may be taken to run from $-|{\bf k}|-\frac{i}{\tau_R}$ to $|{\bf k}|-\frac{i}{\tau_R}$ where $\tau_R=5 \frac{\eta}{s T}$ is the relaxation time in kinetic theory. 
%
It is interesting to note that when increasing $|{\bf k}|$ beyond some critical value $|{\bf k}|_c$, the hydrodynamic poles pass through the logarithmic cut onto the next Riemann sheet, and effectively cease to exist (see Fig.~\ref{fig:two}). Only the non-hydrodynamic branch cut remains, implying that for $|{\bf k}|>|{\bf k}|_c$, hydrodynamics has broken down.

I will summarize the above observations in the form of a 
\begin{lemma}
Given the existence of a local rest frame, hydrodynamics offers a valid and quantitatively reliable description of the energy-momentum tensor even in non-equilibrium situations as long as the contribution from all non-hydrodynamic modes can be neglected.
\end{lemma}
 
Proof: Consider matter possessing a local rest frame everywhere in the space-time patch of interest, such that the local energy density is non-negative in any frame. Pick a convenient global frame (``laboratory frame'') and consider the Fourier decomposition of the energy-momentum tensor in this frame. Now consider real time perturbations $\delta T^{ab}(t,{\bf k})$ around the Fourier zero mode $T^{ab}_{\rm background}$ in the laboratory frame. At some initial time $t_0$, the difference between the local energy-momentum tensor and the background can be viewed as an initial perturbation $S^{ab}(t_0,{\bf k})$. In the limit of small perturbation amplitude $|S^{ab}|\rightarrow 0$, linear response theory applies, cf. Eq.~(\ref{eq:delT}). Furthermore, the retarded two-point function $G_R$ is known to be given by Navier-Stokes hydrodynamics \cite{Kovtun:2012rj} in the small wave-number limit $k\rightarrow 0$. The two-point correlator in Navier-Stokes hydrodynamics possesses hydrodynamic poles (shear and sound poles) in the complex frequency plane. Contour integration as in Eq.~(\ref{eq:delT}) will pick up these poles and lead to a hydrodynamic contribution to $\delta T^{ab}$. As $k$ is increased, the location of the hydrodynamic poles may shift, and they may even disappear from the spectrum completely at some critical wave-number. In addition to the hydrodynamic poles, new, non-hydrodynamic singularities may appear in the complex frequency plane. These non-hydrodynamic singularities, upon contour integration in  Eq.~(\ref{eq:delT}) will lead to a non-hydrodynamic contribution that has to be added to the hydrodynamic part of $\delta T^{ab}$ as in the example given in Eq.~(\ref{eq:delT2}). As the amplitude $S^{ab}$ is increased, other, non-linear structures will contribute to $\delta T^{ab}$ which can be expressed as a sum over integrals of n-point functions with the appropriate powers of the source $S^{ab}$. In the limit of small wave-number, these non-linear corrections to the hydrodynamic part will, upon resummation, shift the hydrodynamic poles locally, and in addition contribute new structures familiar from the hydrodynamic gradient expansion, cf. Eq.~(\ref{Tns}). Away from $k=0$ the non-linear hydrodynamic part of $\delta T^{ab}$ is analytically connected to this familiar structure, giving rise to the generalized hydrodynamic attractor form, until some critical wave-number $k=k_c$ is reached. In addition to the hydrodynamic part, there will be non-linear corrections to the non-hydrodynamic contribution of $\delta T^{ab}$. Thus, the global energy-momentum tensor can be written as $T^{ab}=T^{ab}_{\rm background}+\delta T^{ab}=T^{ab}_{\rm hydro}+T^{ab}_{\rm non-hydro}$. If the non-hydrodynamic contribution can be neglected, the lemma follows trivially.

(Mathematicians probably would want to see a more formal proof than this, so the above lemma should probably be called a conjecture).


The above lemma may seem trivial: once non-hydrodynamic modes are absent, how can the energy-momentum tensor be described by anything else but hydrodynamics? However, when phrased in this fashion, hydrodynamics neither requires equilibrium, nor isotropy, nor infinitesimally small gradients. (It does require the presence of a local rest frame, though \cite{Arnold:2014jva}). The applicability of hydrodynamics is exclusively determined by the relative importance of non-hydrodynamic modes. 

As it stands, the above lemma has at least one potentially important consequence, which is phrased as a
\begin{dilemma}
The phenomenological success of hydrodynamics in describing experimental data from high energy nuclear collisions does not imply near-equilibrium behavior of the matter. Experiments do not directly probe the equilibrium QCD phase diagram at finite $T,\mu$, but explore trajectories in a space with at least one more (non-equilibrium) direction.
\end{dilemma}

\begin{figure}[t]
\includegraphics[width=0.7\linewidth]{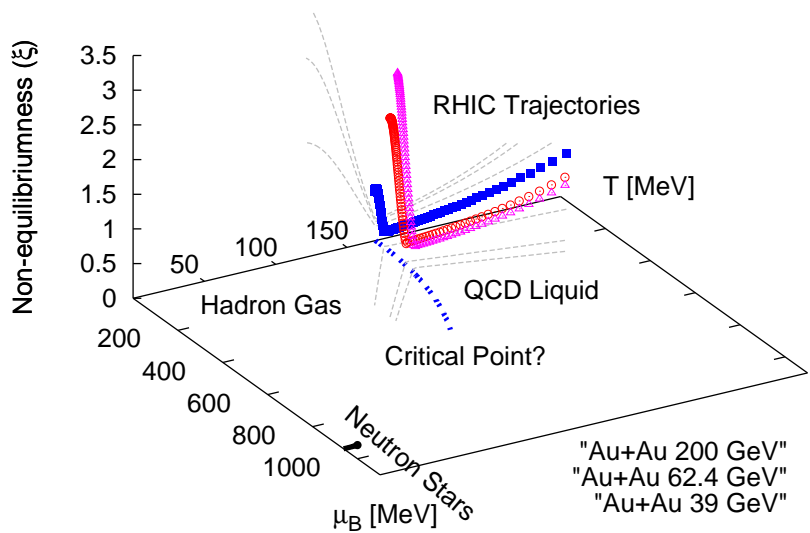}
\caption{A non-artist's calculation of possible RHIC trajectories for Au+Au collisions at various collision energies $\sqrt{s}=39,64,200$ GeV. Rather than displaying the common projections in the temperature - baryon chemical potential ($T,\mu$) plane, trajectories explored by experiment are more likely to explore at least a third non-equilibrium direction (symbols). Projections of the trajectories to the $T,\mu$ and $T,\xi$ planes are indicated as grey dashed lines. A minimum of $\xi\simeq 0.2$ is achieved for $\sqrt{s}=200$ GeV at $T\simeq 0.17$ MeV, which corresponds to $P_L/P_T\simeq 0.86$. For reference, the deconfinement cross-over transition (blue dotted line) and the liquid-gas first order transition (full black line) have been indicated. Note that as the collision energy is lowered, one moves further away from equilibrium.
\label{fig:three}}
\end{figure}

Incomplete equilibration in nuclear collisions is a fact well known to all heavy-ion hydro practitioners in the field and has been pointed out more than a decade ago by Bhalerao, Blaizot, Borghini and Ollitrault in Ref.~\cite{Bhalerao:2005mm}. The above lemma allows me to go one step further since not even near-equilibrium is required for hydrodynamics.  The second part of the dilemma is a direct consequence of the first, yet it probably is not as widely appreciated. I have tried to visualize the last point of the dilemma in Fig.~\ref{fig:three}. To generate the hypothetical trajectories I have matched the pressure anisotropy $P_L/P_T$ to the momentum anisotropy parameter $\xi$, first defined in Ref.~\cite{Romatschke:2003ms}. I use $\xi\, \in[0,\infty)$ to express the degree of non-equilibrium (``non-equilibriumness''), where $\xi=0$ corresponds to the case of local equilibrium. There is an extensive literature in anisotropic hydrodynamics which makes the connection between $P_L/P_T$ and $\xi$ precise
(see e.g. the instructive lecture notes by Strickland, Ref.~\cite{Strickland:2014pga}, Eq.~(3.19)). For the pressure anisotropy itself, I used Navier-Stokes hydrodynamics in Bjorken expansion (cf.~\cite{Keegan:2015avk}) and a viscosity that is given by $\frac{\eta}{s}=\frac{1}{4\pi}$ for $T>0.17$ GeV and rises linearly as the temperature is lowered to reach $\frac{\eta}{s}=1$ at $T=0.1$ GeV (cf.~\cite{Demir:2008tr,Romatschke:2014gna}). The temperature and chemical potential dependence result from a drastically simplified version of Hung and Shuryak, Ref.~\cite{Hung:1997du}, with just one massive degree of freedom in the hadron gas phase, and assuming constant baryon density to entropy ratio. Trajectories are started at $\tau=1$ fm/c with multiplicities representative of experimental measurements \cite{Alver:2010ck} converted to temperature values $T(\tau=1{\rm fm/c})$ as in Ref.~\cite{Habich:2014jna}. Clearly, none of these choices do justice to the much more accurate descriptions that are currently available. However, I expect the sketch in Fig.~\ref{fig:three} to be qualitatively reliable.

\section{Breakdown of Hydrodynamics and Tests}

\begin{figure}[t]
\includegraphics[width=0.7\linewidth]{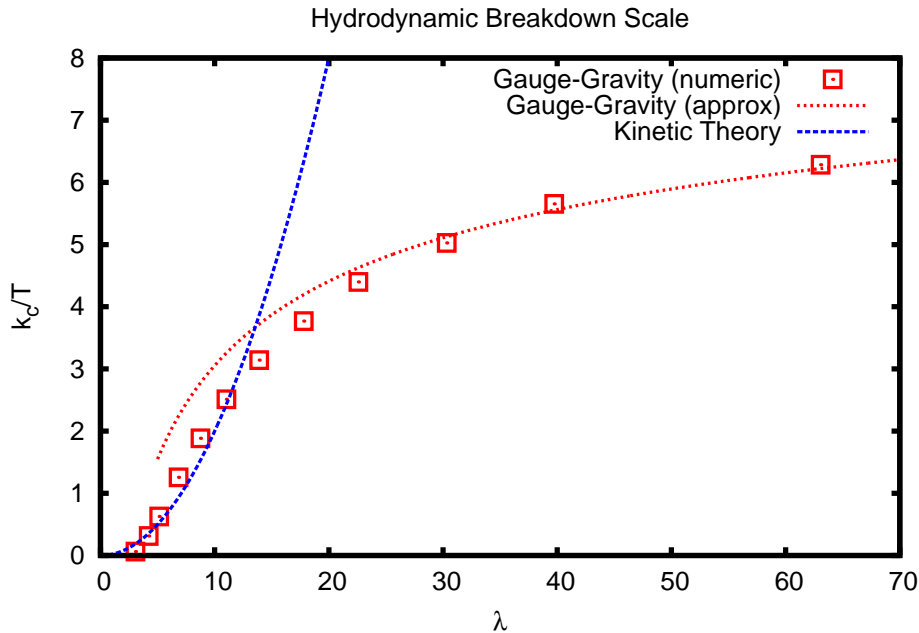}
\caption{Results for the hydrodynamic breakdown scale $k_c$ from weak coupling ('kinetic theory', \cite{Romatschke:2015gic}) and strong coupling ('gauge-gravity', \cite{Grozdanov:2016vgg}) frameworks (the curve labeled 'approx' is a fit to the exact numerical values from Ref.~\cite{Grozdanov:2016vgg}, see text). The regime of applicability for kinetic theory is $\lambda \ll 1$, while for gauge-gravity duality $\lambda \gg 1$ is required. Note that despite the difference in weak-coupling and strong-coupling frameworks, the resulting hydrodynamic breakdown scale is quantitatively similar for moderately strong couplings $\lambda\simeq 10-20$.
\label{fig:fourrep}}
\end{figure}

According to the central lemma in the previous section, hydrodynamics can be used to describe a system if a local rest frame exists and non-hydrodynamic modes are sub-dominant. I have nothing new to say about how to test for the presence of a local rest frame, so I will simply follow everyone else's approach and assume that a local rest frame exists. (This is very likely wrong \cite{Arnold:2014jva} and actually should be thought about more, but I leave this task to dedicated readers).

Contrary to the existence of local rest frames, quite a bit of knowledge now exists on those non-hydrodynamic modes. For the case of kinetic theory with relaxation time $\tau_R$, we know the analytic structure of non-hydrodynamic modes in the two-point function of the energy-momentum tensor, and we know that hydrodynamic modes completely vanish from the spectrum at $|{\bf k}|_c \tau_R \simeq 4.5$ \cite{Romatschke:2015gic}. Using $\tau_R=\frac{5 \eta}{s T}$ and $\frac{\eta}{s}\simeq \frac{5}{g^4}$ for $N=3$ color QCD \cite{Arnold:2003zc}, one finds $|{\bf k}|_c\simeq 0.02 \lambda^2 T$ in terms of the 't Hooft coupling $\lambda\equiv g^2 N$. From gauge gravity duals at large but finite $\lambda$, non-hydrodynamic modes imply the breakdown of hydrodynamics at $|{\bf k}|_c \simeq \frac{\pi T}{4} \ln\left(\frac{6.65 \lambda^{3/2}}{1+105 \lambda^{-3/2}}\right)$ (see Fig.~\ref{fig:fourrep} for exact numerical results from Ref.~\cite{Grozdanov:2016vgg}). These kinetic theory and gauge-gravity results for the hydrodynamic breakdown scale $k_c$ are compared in Fig.~\ref{fig:fourrep}. It is curious to note that -- despite the difference of the kinetic theory and gauge-gravity duality approaches -- the results for $k_c$ turns out to be quantitatively similar in both approaches for moderate values of $\lambda\simeq 10-20$. 

Naively applying the results for $k_c$ to QCD with $\alpha_s\equiv \frac{g^2}{4\pi}\simeq 0.5$ leads to $\frac{k_c(\lambda\simeq 19)}{T}=4-7$, where the higher value is actually coming from the kinetic theory result. Let's be optimistic and take $|{\bf k}|_c \simeq 7 T$. Thus, I claim that no hydrodynamic description is possible for QCD systems smaller than $k_c^{-1}\simeq 0.15$ fm at a typical QCD-scale temperature of $T\simeq 200$ MeV.
The prediction that hydrodynamics \textit{must} break down for $|{\bf k}|^{-1}<0.15$ fm is most likely somewhat useless, because it is very hard to falsify experimentally in nuclear collisions given that the mean proton radius is $0.86$ fm. However, the limit of $0.15$ fm at least constitutes an actual numerical conjecture for the lower bound of the smallest possible droplet of QCD liquid.

It should be pointed out that the results $|{\bf k}|_c\simeq\frac{4.5}{\tau_R}$ and the numerical gauge-gravity results shown in Fig.~\ref{fig:fourrep} are quantitative upper bounds on the domain of hydrodynamic applicability in weak and strong coupling scenarios. However, it is likely that hydrodynamics breaks down before reaching these values of $|{\bf k}|$. One reason why hydrodynamics may break down earlier would be that while hydrodynamic modes do not vanish for $|{\bf k}|<|{\bf k}|_c$, they may become subdominant to certain non-hydrodynamic modes, namely those which happen to be closer to the origin of the complex $\omega$ plane. For ${\cal N}=4$ SYM at $\lambda\rightarrow \infty$, this seems to happen at around $|{\bf k}|\simeq 2\pi T$, which is at around the same value as $k_c$ obtained at finite $\lambda$. Yet another reason why hydrodynamics may break down at scales below $|{\bf k}|_c$ could be non-linear effects. For a particular initial condition, numerical studies of ${\cal N}=4$ at $\lambda\rightarrow \infty$  including full non-linear effects by Chesler \cite{Chesler:2016ceu} seem to indicate a breakdown of hydrodynamics at a scale $|{\bf k}|\simeq T$. These results are fully consistent with the ``most optimistic'' result $k_c\simeq 7 T$ and the resulting hard upper bound for the hydrodynamic breakdown scale, but clearly leave room for sharpening the prediction for $|{\bf k}|_c$.

\begin{figure}[t]
\includegraphics[width=0.45\linewidth]{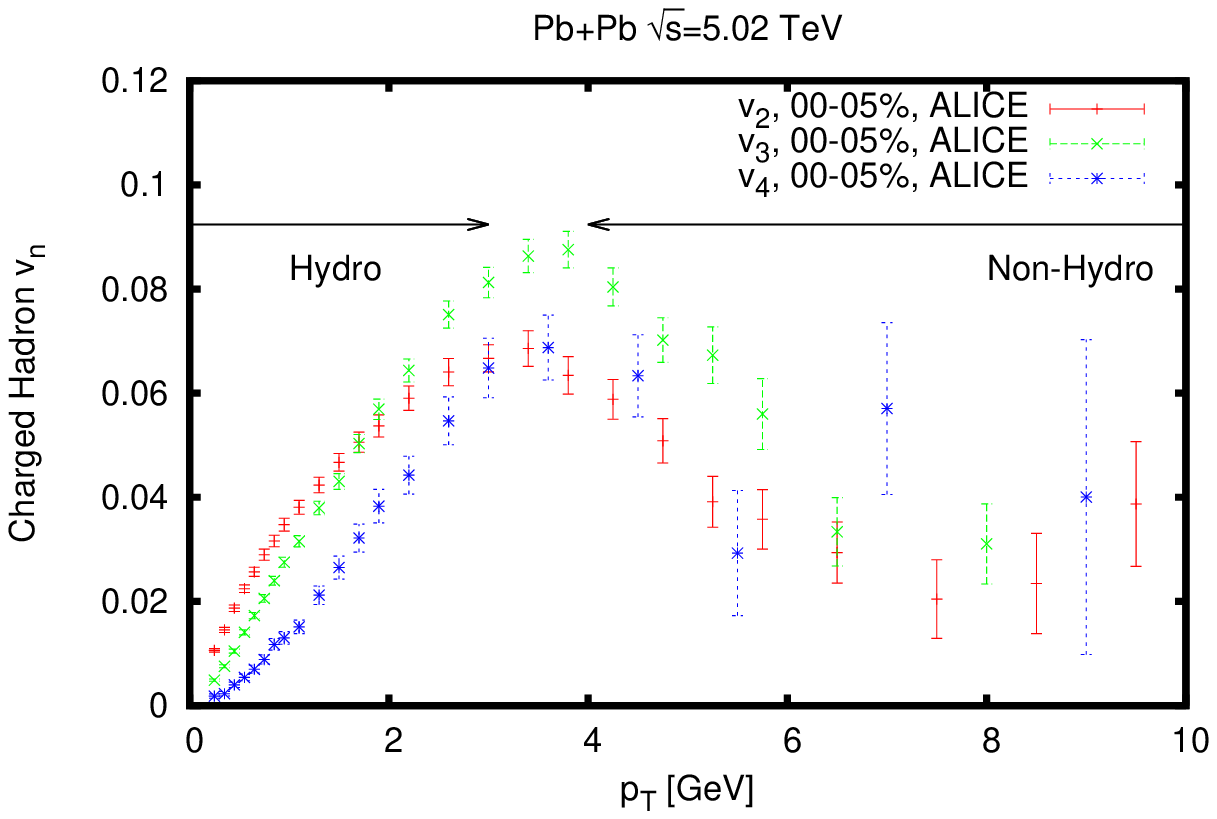}\hfill
\includegraphics[width=0.45\linewidth]{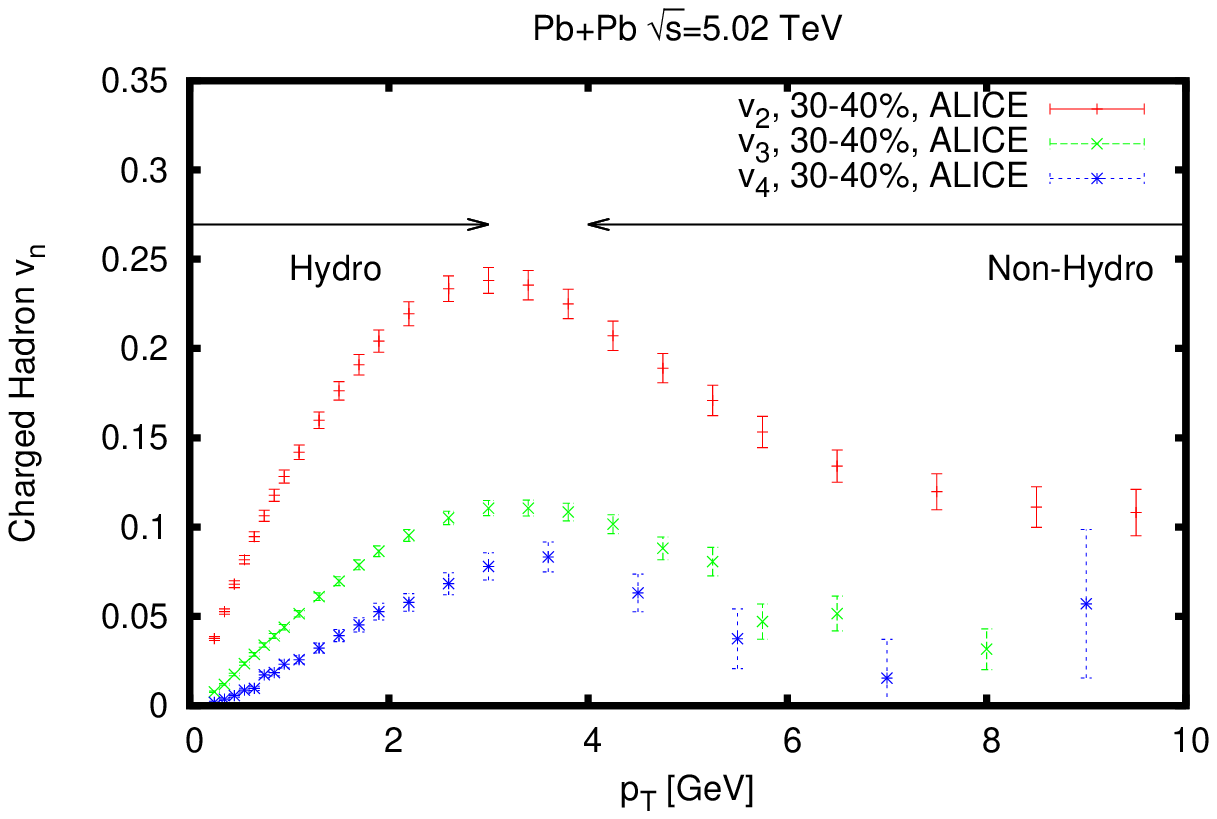}
\caption{Experimental data for flow coefficients $v_n$ as a function of particle $p_T$ for Pb+Pb collisions at $\sqrt{s}=5.02$ TeV (ALICE, \cite{Abelev:2012di}). No hydrodynamic curves are shown, but it is known that hydrodynamics well describes the experimental data in the regime indicated as 'hydro' in the plot \cite{Gale:2012rq}, possibly extending up to $p_T\simeq 3$ GeV. By contrast, for $p_T\gtrsim 4$ GeV, the experimental data seems to deviate systematically from the low-momentum behavior, and I have labeled this region 'non-hydro'. 
\label{fig:five}}
\end{figure}

 A much more direct route to experimentally constrain $k_c$ could be provided by high-momentum data on flow coefficients, see Fig.~\ref{fig:five}. Experimental data for collective flow harmonics in Pb+Pb collisions suggests a change in behavior in the regime between $p_T=3$ GeV to $p_T=4$ GeV. The low momentum region is well described by hydrodynamics \cite{Gale:2012rq}. Assuming that measured particles originated from a constant-temperature freeze-out surface at $T=0.17$ GeV, this would indicate a breakdown of hydrodynamic behavior at $\frac{p_T}{T}=18-23$. In order to relate this scale to the hydrodynamic breakdown scale $k_c\lesssim 7 T$, quantitative calculations of the location of non-hydrodynamic modes in an expanding system are needed.


In the hadronic phase, kinetic theory would predict hydrodynamic modes to dominate for $k<k_c\propto \frac{1}{\tau_R}$, while non-hydrodynamic (particle) modes dominate for $k>k_c$. As the temperature is lowered, $\tau_R\propto \frac{\eta}{s T}$ increases strongly \cite{Demir:2008tr} until $k_c$ falls below the typical system wave-number. From this point onward, most of the system dynamics proceeds according to the non-hydrodynamic particle kinetics, providing a qualitative understanding of the transition from hydrodynamic to particle cascade dynamics (``freeze-out'').

The above statements involve hard lower bounds on the smallest scales at which hydrodynamics applies, and a qualitative understanding of the freeze-out transition. However, a quantitative test of the applicability of hydrodynamics, e.g. through testing sensitivity of results with respect to non-hydrodynamic modes, is desirable in the case of nuclear collisions . Fortunately, the workhorse of relativistic viscous hydrodynamics simulations, ``causal relativistic viscous hydrodynamics'' (which goes by many names and acronyms but is usually associated with the work of M\"uller, Israel and Stewart \cite{Muller:1967zza,Israel:1979wp}) does contain a non-hydrodynamic mode buried within, which may be exploited for testing purposes.
Specifically, besides the usual hydrodynamic modes, the energy-momentum tensor two point function contains a pole located at $\omega_{nh}=-\frac{i}{\tau_\pi}$, where $\tau_\pi$ is the ``viscous relaxation time'' that \emph{also} controls the size of the second order gradient term in the one-point function of $T^{ab}$. Any current numerical hydrodynamics simulation of the matter produced after a relativistic nuclear collision needs a specific value for $\tau_\pi$ as an input. Simulators choose values of $\tau_\pi$ as they see fit, given that the ``correct'' value for $\tau_\pi$ for QCD is not known, and that primary interest is in extracting information about $\frac{\eta}{s},\frac{\zeta}{s}$, not some obscure second-order transport coefficient.

\begin{figure}[t]
\includegraphics[width=0.45\linewidth]{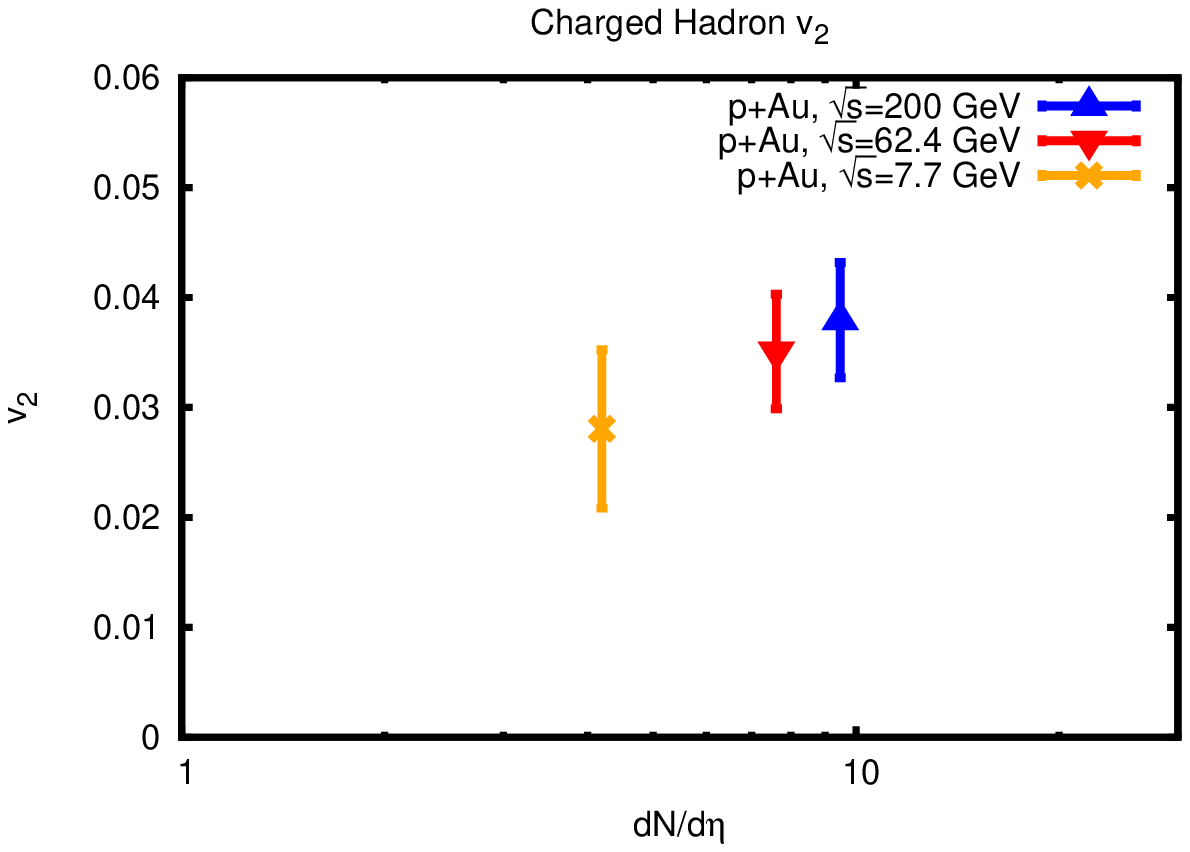}\hfill
\includegraphics[width=0.45\linewidth]{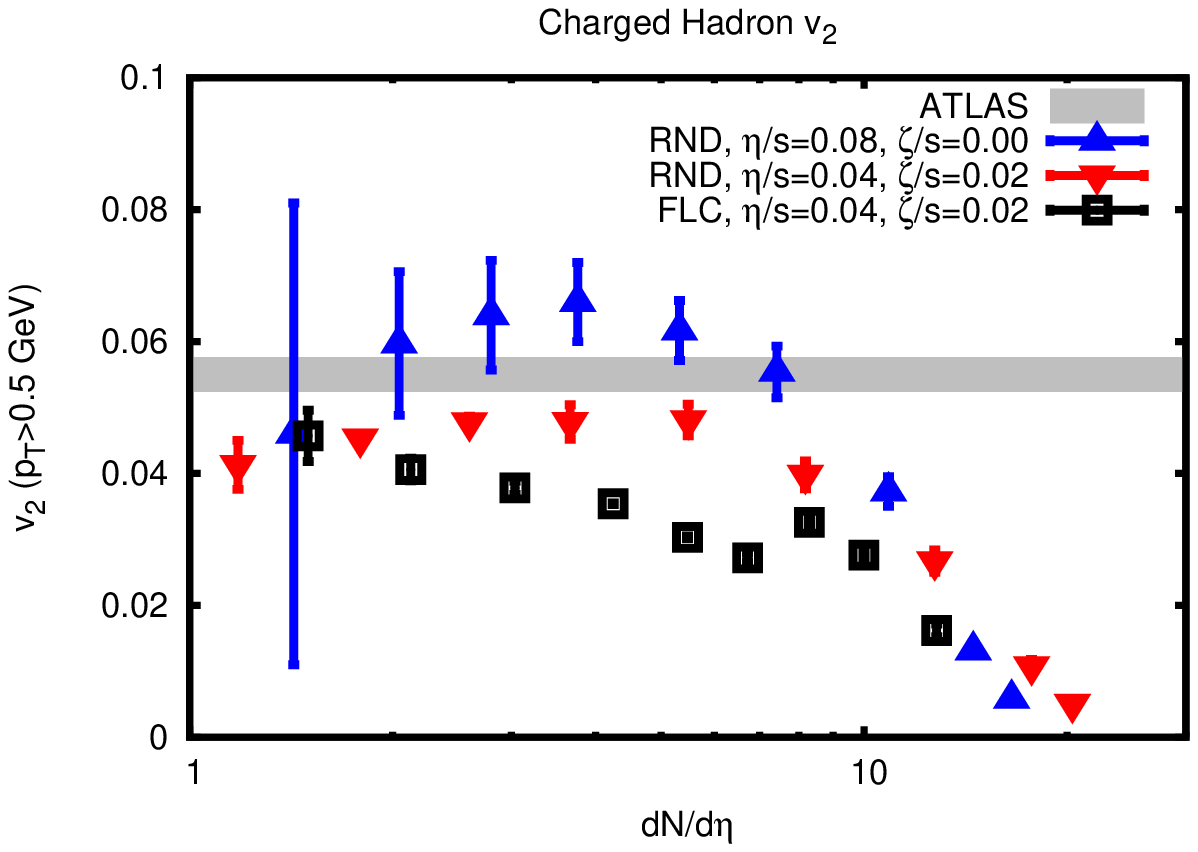}
\caption{Sensitivity of charged hadron $v_2$ on non-hydrodynamic mode for central p+Au collisions (left panel, no $p_T$ cut) and p+p collisions at $\sqrt{s}=7$ TeV (right panel, $p_T$ cut at $0.5$ GeV). Error bars result from varying non-hydrodynamic mode decay time by a factor of two and are found to increase when multiplicity is lowered. In the case of p+p collisions, this indicates that hydrodynamics is not applicable for $\frac{dN}{d\eta}<2$. Figures based on results from Ref.~\cite{Romatschke:2015gxa} and from Ref.~\cite{Habich:2015rtj}, respectively.
\label{fig:four}}
\end{figure}

However, varying $\tau_\pi$ around some ``fiducial'' value does vary the decay-time of the non-hydrodynamic mode inherent to causal relativistic hydrodynamics, thus offering a direct handle on the sensitivity of final results on the non-hydrodynamic mode. This can be implemented in practice in relativistic viscous hydrodynamic simulations by running simulations at multiple values of $\tau_\pi$ and expressing final results in terms of a mean value and a systematic error bar covering the variations of final results from changing $\tau_\pi$. Examples are shown in Fig.~\ref{fig:four} for the case of central p+Au collisions at various values of $\sqrt{s}$ and p+p collisions at $\sqrt{s}=7$ TeV. While the sensitivity on non-hydrodynamic modes is not vanishingly small, the error bars do seem to signal the applicability of hydrodynamics to both p+Au and p+p collisions in general. However, in the case of p+p collisions and $\frac{dN}{dY}<2$, the error bars become large, signaling strong sensitivity of the result to non-hydrodynamic modes. This empirical result seems to indicate that hydrodynamics breaks down in p+p collisions for $\frac{dN}{dY}<2$. This multiplicity value of the hydrodynamic breakdown corresponds well to the results derived by Spalinski \cite{Spalinski:2016fnj}. It would be interesting to repeat these sensitivity tests for hydrodynamics with a different non-hydrodynamic mode structure, for instance along the lines suggested in Ref.~\cite{Heller:2014wfa}.

\section{Concluding Remarks}

\begin{enumerate}
\item
I would argue that there is hard experimental evidence, e.g. through the phenomenon of jet modifications, for the presence of strongly interacting QCD matter created in nuclear collisions.  As argued in this note, I am doubtful about the hard evidence for this matter to be equilibrated.
\item
The physics of non-hydrodynamic modes is a rich and a barely studied subject. Given that non-hydrodynamic modes play an important role in the applicability and breakdown of a hydrodynamic descriptions, I believe those non-hydro modes should receive more attention, from theorists and experimentalists alike.
\item
The central lemma in section \ref{sec:this} also applies to the case of diffusion, not only momentum transport. In particular, this implies that a constitutive equation of the form $J=\sigma E$ could hold in the early-time, out-of-equilibrium regime following a nuclear collision, if non-hydro modes are sub-dominant. This could potentially explain a longer-than-expected life-time of the magnetic field which is critical to experimental detection of the Chiral Magnetic Effect \cite{Skokov:2016yrj} (see also Ref.~\cite{Lin:2013sga}).
\item
As outlined in the central dilemma in section \ref{sec:this}, the experimental search for the QCD critical point will necessarily explore trajectories in some non-equilibrium space (cf. Fig.~\ref{fig:three}). This implies that the standard equilibrium theory of critical fluctuations strictly speaking does not apply, and one should try to understand non-equilibrium effects (see e.g. Ref.~\cite{Mukherjee:2015swa}) in order to correctly interpret the experimental data.
\item
In view of the 'QGP drop size lower bound' of $0.15$ fm, it is maybe not surprising that the matter created in p+p collisions would behave hydrodynamically. At this scale, however, p+p collisions may not be the ultimate drop size test. QCD-QED couplings allow fluctuations of electrons to e.g. quark pairs, thus opening up the possibility of local energy deposition reminiscent of p+p collisions occurring in $e^+$-$e^-$ collisions (cf. Refs.~\cite{Hoang1987,Becattini:2001fg,Ferroni:2011fh}). Data on $e^+$-$e^-$ collisions taken at e.g. LEP should be re-analyzed with modern tools in order to find (or rule out) hydrodynamic behavior in these systems.
\item
The fact that experimental data shows a qualitative change in trend from hydrodynamic behavior at low momenta to non-hydrodynamic behavior at high momenta suggests a potential experimental handle on the hydrodynamic breakdown scale $k_c$ in QCD. This potential connection should be made quantitative in further studies.
\item
The entire discussion in this note ignores the presence of hydrodynamic thermal fluctuations, which arise in $SU(N)$ gauge theories at any finite number $N$. The subfield of relativistic hydrodynamics with thermal fluctuations is still in its infancy, but potentially can have important phenomenological consequences \cite{Kovtun:2003vj,Kovtun:2011np,PeraltaRamos:2011es,Kovtun:2012rj,Young:2013fka,Murase:2013tma,Crossley:2015evo,Nahrgang:2016ayr,Akamatsu:2016llw}.
\item
A recurring problem of non-standard cosmology (so-called ``viscous cosmology'', cf.~\cite{Zimdahl:2000zm,Floerchinger:2014jsa}) seems to be that ``interesting'' deviations from standard cosmology occur when gradient corrections become order unity. In the ``old-fashioned'' picture of hydrodynamics, this was not acceptable since order unity corrections heralded the breakdown of applicability of the theory. In view of the central lemma in section \ref{sec:this}, it could be interesting to determine the relevant non-hydrodynamic modes in cosmology and re-evaluate the regime of applicability of viscous cosmologies.
\end{enumerate}
 
\section{Acknowledgments}

I am indebted to many colleagues for countless fruitful discussions on the topics mentioned in this note. In particular I would like to thank J.~Cassalderrey-Solana, T.~DeGrand M.~Floris, J.~Nagle, A.~Kurkela, J.~Schukraft, M.~Spalinski, A.~Starinets, M.~Stephanov and W.~van der Schee for discussions, and the organizers of the CERN-theory workshop ``The Big Bang and the little bangs'', the Oxford workshop on ``Non-Equilibrium Physics and Holography'', the Santiago de Compostela meeting on ``Numerical Relativity and Holography'' and the Technion workshop on ``Numerical Methods for AdS spaces'' for providing interesting meetings and many stimulating discussions. Moreover, I would like to thank F.~Becattini, M.~Heller, S.~Morwczynski, J.~Nagle (again!), J.~Noronha, J.~Schukraft (again!), L.~Yaffe and B.~Zajc for their detailed and encouraging comments on the first version of this note and S.~Grozdanov, N.~Kaplis and A.~Starinets for providing their exact numerical data on $k_c$ from Ref.~\cite{Grozdanov:2016vgg}. I would like to thank the U.S. Department of Energy for providing financial support under DOE award No. DE-SC0008132. Finally, I am grateful to Lufthansa for many hours without phone or internet, leaving me no other choice than to think about physics. 

\bibliography{noneq}

\end{document}